# Learning about Potential Users of Collaborative Information Retrieval Systems

**Madhu Reddy, Ph.D.**
College of Information Sciences and Technology
The Pennsylvania State University
University Park, PA
mreddy@ist.psu.edu

**Bernard J. Jansen Ph.D.**
College of Information Sciences and Technology
The Pennsylvania State University
University Park, PA
jjansen@ist.psu.edu

## ABSTRACT
One of the key components of designing usable and useful collaborative information retrieval systems is to understand the needs of the users of these systems. Our research team has been exploring collaborative information behavior in a variety of organizational settings. Our research goals have been two-fold: First, to develop a conceptual understanding of collaborative information behavior and second, gather requirements for the design of collaborative information retrieval systems. In this paper, we present a brief overview of our fieldwork in a three different organizational settings, discuss our methodology for collecting data on collaborative information behavior, and highlight some lessons that we are learning about potential users of collaborative information retrieval systems in these domains.

## INTRODUCTION
One of the key components of designing usable and useful collaborative information retrieval (CIR) systems is to understand the needs of the users of these systems. However, most current models and studies of human information behavior have focused primarily on individual needs and practices. For example,

- Kuhlthau's studies (1989; 1989; 1991) of high school students examined individual information seeking behavior; therefore, her model conceptualized information seeking as an individual activity.

- Ellis' model reflects his studies' (1993; 1997) emphasis on information seeking as an individual activity.

- Wilson (1981) developed his model after examining information needs and seeking (IN&S) user studies. The model is his conceptualization of the IN&S process, but it also reflects the individual nature of the information seeking typified in earlier user studies.

- Leckie, Pettigrew and Sylvain.'s model (1996) was developed from a literature survey of studies examining the largely individual information seeking behavior of engineers, physicians, and lawyers.

These studies focused on the individual primarily because information seeking, and the related area of information searching and retrieval (IS&R), was viewed as being embedded in individual not collaborative work context.

A number of studies examining information seeking in a wide variety of collaborative settings (Fidel et al., 2000; Foster, 2006) are now challenging this perspective. These studies are starting to pave the way for both a more detailed conceptual understanding of collaborative information seeking and the improved design of CIR systems.

Our research team has been exploring collaborative information seeking practices in a variety of organizational settings for a number of years. We have used the term *collaborative information behavior* (CIB) in our research studies (Reddy and Jansen, 2008). Our research goals have been two-fold: First, to develop a conceptual understanding of CIB and second, gather requirements for the design of CIR systems.

In this workshop paper, we focus our attention on the empirical fieldwork aspect of our research and what we are learning about the potential users of CIR systems.

In the remainder of the paper, we present a general overview of our fieldwork in a three different organizational settings, discuss our methodology for collecting data on CIB, and some lessons that we are learning about users.

## RESEARCH SITES

We have been conducting research into CIB in three different organizational settings in two major domains. Each of these settings provides a unique opportunity to examine synchronous IS&R activities within a collaborative environment.

- Patient Care Teams (Health Care) – We have conducted studies examining CIB in the surgical intensive care unit and emergency departments of two hospitals. (Reddy et al., 2002; Reddy and Spence, 2006) A hospital with its numerous information resources and focus on collaborative patient care teams is a natural setting for investigating CIB. These units are, arguably, the most information-intensive and collaborative settings in the hospital. In these dynamic and fast-paced environments, team members face numerous challenges to finding needed information such as fragmented information resources. Furthermore, team members cannot afford to make mistakes; therefore, they must ensure that the information shared among team members is accurate. Because of these challenges, collaboration during information seeking activities is an integral aspect of the work in the unit.

- Information Technology (IT) Teams (Health Care) - We are currently conducting a study a rural, regional hospital's IT teams CIB practices. This setting is particularly suited for this research. First, collaboration is an integral part of an IT team's work. Whether it is resolving a help desk issue, developing an application for hospital staff, or implementing some software/hardware component, team members constantly interact with each other to accomplish their tasks. Second, the rural nature of the hospital makes it difficult to recruit "experts" with specialized skills. Hence, there is an added emphasize on team members working together to solve a problem. Finally, the IT teams of this hospital not only support their own hospitals' staff but also the IT needs of three smaller hospitals in the region. Therefore, there is a great deal of collaboration between the IT staff in these different hospitals, which introduces the distributed aspect inherent in many CIB efforts.

- Student Teams (Education) – We are conducing a study of student teams and their CIB practices in a junior level course in a technology based major. Understanding CIB is of increasing importance in this domain. As educational institutions develop knowledge workers in information technology areas, they are increasingly placing students in teams that bridge departments, campuses, and even institutions. Within this collaborative and team environment, the classroom provides a complex CIB environment that is important to investigate and is an excellent laboratory for exploring aspects of CIB.

Interestingly given the information intensive nature of each of these domains, none of these research sites utilized any explicit CIR systems.

## METHODS

In our field research, we have been utilizing qualitative methods (Reddy and Spence, 2008) and quantitative methods such, as surveys, to examine CIB (Spence et al., 2005). In fact, we are using surveys to examine student teams. One of our goals is to develop a standard survey instrument that we can use across domains for a wide variety of teams.

Yet, our primary empirical approach is ethnographic fieldwork. Ethnographic observation is designed to provide a deep understanding and support rich analytical description of a phenomenon, as part of an iterative cycle of observation and analysis (Strauss and Corbin, 1990). It seeks not just to document actions, but to examine what is experienced in the course of these actions.

Studying CIB requires careful observation and questioning. Multiple people need to be interviewed, and only with sufficient observation can we identify different CIB practices and their effects on daily work activities. Since people often cannot tell a researcher what they actually do in practice (rather than what they are supposed to do), it has been found more useful to both interview and observe study participants. For instance, in an example of tacit understandings, people may tell a researcher that they "officially" ask the unit pharmacist when seeking information about a particular medication. However, in practice, they may be observed to bypass the unit pharmacist and directly ask a pharmacist outside the unit about the medication.

It is probable that many other tacit understandings about how people collaborate when seeking information exist (e.g., assumptions about the quality of the information, background of individuals, individual's knowledge); only a field study can reveal these tacit understandings. Indeed, only a field study can uncover CIB practices, can tell us what issues are important for which groups of people, and most importantly, can tell us *why* these issues are important. From these findings, one can tailor other methods to shed greater insight on aspects of CIB.

## LESSONS LEARNED

Through our fieldwork, we are beginning to learn a number of interesting lessons about potential users of CIR systems. We highlight some of the lessons below.

- Communication is key for synchronous CIB – Communication is essential for successful collaboration. This is especially true when searching for information. Team members continuously exchanged information about the *search process* as they collaborated during information seeking and retrieval activities. Exchanging information about the search process allowed team members in our studies to stay on track and alerted other team members when they may be taking a wrong search path. An understanding of the search process also provided validity to the information.

- Targeted vs. general information search – Team members most often collaborated to find information in order to answer specific questions. The information seeking was targeted and specific. This is not to say that team members knew what they were looking for (or where to look for it). However, this highlights the issue of the role that collaboration in information retrieval may play in organizational settings vs. other environments. We found that the collaboration in these settings was often not for general knowledge acquisition but rather for specific purposes (e.g., address an explicit need).

- Electronic and Non-electronic sources – Information in these organizational settings (especially in the healthcare domain) is scattered across a number of different resources including electronic, paper, and human. Therefore, it was not a simple task to find the source of information. This required utilizing a variety of resources that may not be simple to capture in an electronic format. So, in a CIR process, there may be information that is not available electronically. This raises the issue of how we can support these activities during the CIR process.

Other contextual, cognitive, and technological findings are presented in (Reddy and Jansen, 2008).

## CONCLUSION

Through our research, we are starting to understand CIB across different settings. We have, for instance, identified some work features that trigger CIB (Reddy and Spence, 2008). We have also identified features through the fieldwork that we believe are essential to CIR systems; in particular, awareness and communication. We are exploring these features in our CIR prototypes (Reddy and Jansen, 2008).

Through the workshop, we hope to learn more about how we can better utilize our field research in the CIR system design and issues about CIR systems that we can explore in our fieldwork. We also believe that the results from our fieldwork can help inform others involved in the design of CIR systems.

## REFERENCES

Ellis, D., D. Cox and K. Hall (1993). "A comparison of the information seeking patterns of researchers in the physical and social sciences." *Journal of Documentation* 49(4): 356-369.

Ellis, D. and M. Haugan (1997). "Modeling the Information Seeking Patterns of Engineers and Research Scientists in an Industrial Environment." *The Journal of Documentation* 53(4): 384-403.

Fidel, R., H. Bruce, A. M. Pejtersen, S. Dumais, J. Grudin and S. Poltrock (2000). "Collaborative Information Retrieval (CIR)." *New Review of Information Behaviour Research: Studies of Information Seeking in Context* 1(1): 235-247.

Foster, J. (2006). Collaborative Information Seeking and Retrieval. Annual Review of Information Science and Technology. B. Cronin. Medford, NJ, Information Today, Inc. 40**:** 329-356.

Kuhlthau, C. C. (1989). "The information search process of high-middle-low achieving high school seniors." *School Library Media Quarterly* 17: 224-228.

Kuhlthau, C. C. (1989). "The information search process of high-middle-low achieving high school seniors." *School Library Media Quarterly* 17: 224-228.

Kuhlthau, C. C. (1991). "Inside the Search Process: Information Seeking from the User's Perspective." *Journal of the American Society for Information Science* 42(5): 361-371.

Leckie, G. J., K. E. Pettigrew and C. Sylvain (1996). "Modeling the Information Seeking of Professionals: A General Model Derived from Research on Engineers, Health Care Professionals, and Lawyers." *Library Quarterly* 66(2): 161-193.

Reddy, M. and J. Jansen (2008). "A Model for Understanding Collaborative Information Behavior in Context: A Study of Two Healthcare Teams." *Information Processing and Management* 44(1): 256-273.

Reddy, M., W. Pratt, P. Dourish and M. Shabot (2002). Asking Questions: Information Needs in a Surgical Intensive Care Unit. of the American Medical Informatics Association Fall Symposium (AMIA'02), San Antonio, TX.: 651-655.

Reddy, M. and P. R. Spence (2006). Finding Answers: Information Needs of a Mulitdisciplinary Patient Care Team in an Emergency Department In Proc. of American Medical Informatics Association Fall Symposium (AMIA'06). Washington, DC**:** 649-653.

Reddy, M. and P. R. Spence (2008). "Collaborative Information Seeking: A field study of a multidisciplinary patient care team " *Information Processing and Management* 44(1): 242-255.

Spence, P. R., M. Reddy and R. Hall (2005). A Survey of Collaborative Information Seeking of Academic Researchers. the ACM Conf. on Supporting Group Work (GROUP'05), Sanibel Island, FL.: 85-88.

Strauss, A. and J. Corbin (1990). Basics of Qualitative Research: Grounded Theory Procedures and Techniques. Newbury Park, CA, Sage Publications.

Wilson, T. D. (1981). "On User Studies and Information Needs." *Journal of Documentation* 37(1): 3-15.
- 4 -